\documentclass[12pt,dvips]{article}
\usepackage[dvips]{graphicx}
\usepackage{graphicx}

\graphicspath{{plots/}}
\DeclareGraphicsExtensions{.eps.gz,.eps,.ps,.ps.gz}

\parskip=\itemsep               
\newcommand{\lwig}{\mbox{\,\raisebox{.3ex}
    {$<$}$\!\!\!\!\!$\raisebox{-.9ex}{$\sim$}\,}}
\newcommand{\gwig}{\mbox{\,\raisebox{.3ex}
    {$>$}$\!\!\!\!\!$\raisebox{-.9ex}{$\sim$}}\,}
\newcommand{\lambdabar}{{\hbox{$\lambda_e$\kern-1.9ex\raise+0.45ex\hbox{--}
\kern+0.2ex}}}

\date{\empty}

\title{{\normalsize\rightline{DESY 02-001}\rightline{hep-ph/0201139}}
\vskip 1cm 
\bf Black Holes at Neutrino Telescopes
       \vspace{21mm}} 
\author{
M. Kowalski\\[4mm]
Deutsches Elektronen-Synchrotron DESY, Zeuthen, Germany\\[4mm]
A. Ringwald, H. Tu\\[4mm] 
Deutsches Elektronen-Synchrotron DESY, Hamburg, Germany}

\begin{document}
\begin{titlepage} 
  \maketitle
\begin{abstract}
In scenarios with extra dimensions and TeV-scale quantum gravity,
black holes are expected to be produced in the collision of light particles 
at center-of-mass energies above the fundamental Planck scale with small impact parameters.
Black hole production and evaporation may thus be studied in detail at the  
Large Hadron Collider (LHC).
But even before the LHC starts operating, 
neutrino telescopes such as  AMANDA/IceCube,  ANTARES,  
Baikal, and RICE have an opportunity to search for 
black hole signatures. Black hole production in the scattering of 
ultrahigh energy cosmic neutrinos on nucleons in the ice or water  may initiate 
cascades and through-going muons with distinct characteristics above the Standard Model rate. 
In this Letter, we investigate the sensitivity of neutrino telescopes  
to black hole production and compare it to the one expected at the Pierre Auger Observatory, 
an air shower array currently under construction, and at the LHC. 
We find that, already with the currently available data, AMANDA and RICE should be able
to place sensible constraints in black hole production parameter space, which are 
competitive with the present ones from the air shower facilities Fly's Eye and AGASA.
In the optimistic case that a ultrahigh energy cosmic neutrino flux significantly higher than the 
one expected from cosmic ray interactions with the cosmic microwave background radiation
is realized in nature, one even has discovery potential for black holes at neutrino telescopes 
beyond the reach of LHC.     
\end{abstract}


\thispagestyle{empty}
\end{titlepage}
\newpage \setcounter{page}{2}

{\em 1.}
The possibility that black holes will be produced in the 
collision of two light particles at center-of-mass (cm) energies above the Planck scale
with small impact parameters~\cite{'tHooft:1987rb}  
seems now within reach  in the context of theories with
$\delta = D-4\geq 1$ large compact~\cite{Arkani-Hamed:1998rs} 
or warped~\cite{Randall:1999ee} extra dimensions 
and a low fundamental Planck scale 
$M_D\,\gwig$ 1 TeV characterizing quantum 
gravity.
In these theories one might expect the copious production of black holes 
in high energy collisions
at cm energies above 
$M_D$~\cite{Argyres:1998qn}.
Correspondingly, the Large Hadron Collider (LHC)~\cite{Evans:2001mn}, 
expected  to have a first physics run in  2006, may turn into a factory of black holes
at which their production and evaporation may be studied in 
detail~\cite{Giddings:2001bu,Dimopoulos:2001hw,Dimopoulos:2001qe,Hossenfelder:2001dn}. 

Black hole production and subsequent decay in the scattering of ultrahigh energy cosmic neutrinos
on nucleons in the atmosphere may initiate quasi-horizontal air showers 
far above the Standard Model rate~\cite{Feng:2002ib,Emparan:2001kf} and with distinct 
characteristics~\cite{Anchordoqui:2001ei}. The reach 
of the Pierre Auger Observatory~\cite{Zavrtanik:2000zi} for extensive air showers, 
expected to be completed by the end of 2003,
in the black hole production parameter 
range has been thoroughly investigated in Refs.~\cite{Feng:2002ib,Ringwald:2001vk,Anchordoqui:2001cg}.  
It was found that, depending on the size of the flux of ultrahigh energy cosmic neutrinos, 
the first signs of black hole production may be observed at Auger already before
the start of the LHC. Furthermore, it was shown that already now sensible constraints
on black hole production can be obtained from the non-observation of horizontal 
showers~\cite{Baltrusaitis:1985mt,Yoshida:2001}
by the Fly's Eye collaboration~\cite{Baltrusaitis:1985mx} and the 
Akeno Giant Air Shower Array (AGASA) collaboration~\cite{Chiba:1992nf}, 
respectively.  These constraints turn out to be 
competitive with other currently available constraints on TeV-scale gravity which are mainly based on interactions associated with
Kaluza-Klein gravitons, 
according to which a fundamental Planck scale as low as $M_D = {\mathcal O}(1)$ TeV is still allowed for $\delta\geq 4$ flat 
or $\delta\geq 1$ warped extra dimensions~\cite{Peskin:2000ti}\footnote{For an exhaustive list of references in this 
context, see also Ref.~\cite{Ringwald:2001vk}.}. 

In this Letter, we want to consider the sensitivity of neutrino telescopes such as 
AMANDA/Ice\-Cube~\cite{Halzen:1999jy}, 
ANTARES~\cite{Montaruli:2002gh},  Baikal~\cite{Domogatsky:2001br}, NESTOR~\cite{Grieder:2001kr}, 
and RICE~\cite{Kravchenko:2001id}
to black hole production in the scattering of ultrahigh energy cosmic neutrinos on 
nucleons in ice or water, considerably extending  a first exploratory
study~\cite{Uehara:2001yk}.
After briefly reviewing the phenomenological model of black hole 
production adopted, we calculate the rate expectation for events contained in the 
fiducial volume of a neutrino telescope.   
We then extend our calculations to rates from through-going muons.
Finally, we discuss the achievable sensitivity for detection of 
black hole production and compare it with that from air-shower 
facilities and the LHC.

{\em 2.}
Our phenomenological model of black hole production and decay will be based on the 
following working hypothesis which summarizes the current understanding of the 
production and decay of black holes in TeV-scale gravity 
scenarios~\cite{Argyres:1998qn,Emparan:2000rs,Giddings:2000ay,Giddings:2001bu,Dimopoulos:2001hw,%
Dimopoulos:2001qe,Hossenfelder:2001dn}.

It is assumed that 
at trans-Planckian parton-parton cm energies squared, $\hat s\gg M_D^2$,  
and at small parton-parton impact parameters, $b\ll r_S\,(M_{\rm bh}=\sqrt{\hat s})$, i.\,e. 
at impact parameters much smaller than the Schwarzschild radius $r_S$ of a $(4+\delta )$-dimensional black 
hole with mass $M_{\rm bh}=\sqrt{\hat s}$~\cite{Myers:1986un}\footnote{We define $M_D$ as in 
Ref.~\cite{Giudice:1999ck}. 
Equation~(\ref{schwarzsch}) is valid as long as $r_S\ll R_c$, with $R_c$ being the compactification 
or curvature radii in the flat or warped scenario, respectively.}, 
\begin{equation}
\label{schwarzsch}
r_S =\frac{1}{M_D}
\left[
\frac{M_{\rm bh}}{M_D}
\left(
\frac{2^\delta \pi^{\frac{\delta -3}{2}}\,\Gamma\left( \frac{3+\delta}{2}\right)}{2+\delta}
\right)
\right]^{\frac{1}{1+\delta}}
\,,
\end{equation} 
a black hole forms with a cross section
\begin{equation}
\label{sig_bh_geom}
\hat\sigma (ij\to {\rm bh})\equiv  
\hat\sigma^{\rm bh}_{ij} (\hat s ) \approx \pi\,r_S^2 
\left( M_{\rm bh}=\sqrt{\hat s} \right)\,
\theta\left( \sqrt{\hat s} -M_{\rm bh}^{\rm min}\right) 
\,.
\end{equation}
Here, $M_{\rm bh}^{\rm min}\gg M_D$ parametrizes the cm energy above which the
semiclassical reasoning mentioned above is assumed to be valid.
The contribution of black hole production to the neutrino-nucleon  cross section is then
obtained as the convolution 
\begin{equation}
\label{sig_nuN_bh_pdf}
\sigma_{\nu N}^{\rm bh} (s) =
\sum_i \int^1_0
{\rm d}x\,f_i (x,\mu )\,\hat \sigma_{\nu i}^{\rm bh} (xs)
\, .
\end{equation}
Here, $s=2\,m_N\,E_\nu$, with $m_N=(m_p+m_n)/2$ denoting an isoscalar nucleon mass and $E_\nu$
the neutrino energy in the laboratory frame, denotes the 
neutrino-nucleon cm  energy squared. The sum 
extends over all partons in the nucleon, with parton distribution functions $f_i(x,\mu )$ and 
factorization scale $\mu$. As in Ref.~\cite{Ringwald:2001vk}, we have used in the calculation of the
cross section~(\ref{sig_nuN_bh_pdf}) various sets of parton 
distributions as they are implemented in the parton distribution library 
PDFLIB~\cite{Plothow-Besch:1993qj}. In the following we display only the results obtained with 
the CTEQ3D~\cite{Lai:1995bb} parton distributions with $\mu = {\rm min}\,(\sqrt{\hat s},10\ {\rm TeV})$, 
keeping in mind that uncertainties associated with different parton distribution sets 
are in the ${\mathcal O}(20)\, \%$ range, whereas a choice of the other natural factorization scale 
$\mu = r_S^{-1}$~\cite{Giddings:2001bu,Emparan:2001kf} decreases the predicted rates
by a factor of ${\mathcal O}(2)$.

It is fair to say that there remain nevertheless substantial uncertainties in the size of the cross section.
As noted in Ref.~\cite{Voloshin:2001vs}, the semiclassical production of black
holes resembles largely the problem of baryon and lepton number violating processes 
(``sphaleron~\cite{Klinkhamer:1984di} production'') in multi-TeV ($\sqrt{\hat s}\gg m_W/\alpha_W$) 
particle collisions in the standard electroweak theory~\cite{Ringwald:1990ee,Morris:1991bb,Gibbs:1995cw} 
and more generally the problem of multi-particle production in weakly coupled theories~\cite{Cornwall:1990hh}, 
which is not yet completely understood~\cite{Mattis:1992bj}.  
There might be an additional exponential suppression factor rendering semiclassical 
sphaleron, multi-particle or black hole production unobservable in the TeV range~\cite{Voloshin:2001vs} 
(see, however, Ref.~\cite{Dimopoulos:2001qe}). 
Nevertheless, in this Letter we shall confine ourselves  to the simple geometric cross section 
estimate~(\ref{sig_nuN_bh_pdf}) and postpone an inclusion of a possible exponential suppression factor 
to a longer write-up~\cite{Kowalski:inprep}. 

\begin{figure}
\begin{center}
\includegraphics*[bbllx=20pt,bblly=221pt,bburx=570pt,bbury=608pt,%
width=12cm]{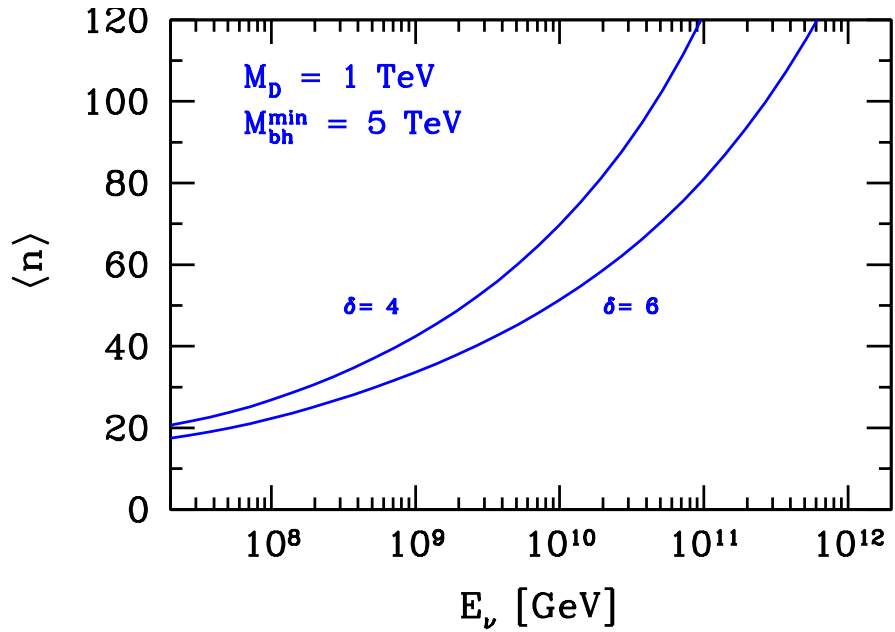}
\caption[dum]{\label{multiplicity}
The average number $\langle n\rangle$ of Standard Model particles~(\ref{eq:av_mult}), into which a 
black hole produced in neutrino-nucleon scattering decays, as a function of the
incident neutrino energy in the laboratory frame, $E_\nu$. 
}
\end{center}
\end{figure}

After production, the black holes decay primarily 
via Hawking radiation~\cite{Hawking:1974rv}, corresponding to a temperature 
\begin{equation}
\label{eq:hawk_temp}
T_{\rm H}= \frac{\delta +1}{4\pi r_S}\,,
\end{equation}
into a large number~\cite{Dimopoulos:2001hw},   
\begin{equation}
\label{eq:av_mult}
\langle n\rangle \approx \frac{1}{2}\,\frac{M_{\rm bh}}{T_{\rm H}}\,,
\end{equation}
of hard quanta (cf. Fig.~\ref{multiplicity}), with energies approaching several hundreds of GeV in the cm system. 
We shall assume that the average multiplicity~(\ref{eq:av_mult}) refers to 
Standard Model particles~\cite{Emparan:2000rs} -- an assumption which is still under
debate~\cite{Casadio:2001wh} --  in a ``flavor"-democratic fashion.

{\em 3.}
Let us consider now the rate of black hole events taking place within the fiducial volume of a neutrino telescope. 
The number of black hole events initiated by neutrino-nucleon scattering occurring per unit time $t$ 
and solid angle $\Omega$ with 
energy larger than a threshold energy $E_{\rm th}$ inside a subsurface detector volume 
$V$ is (see e.g. Ref.~\cite{Morris:1991bb}) 
\begin{equation}
\label{rate_contained}
\frac{{\rm d}^2 N_{\rm casc}^{\rm bh}}{{\rm d}t\,{\rm d}\Omega} 
= \frac{\rho\,V}{m_N}
\int^\infty_{E_{\rm th}} {\rm d}E_\nu\,
F_\nu (E_\nu )\,\sigma_{\nu N}^{\rm bh}(E_\nu )\,
\exp\left[ -\sigma^{\rm tot}_{\nu N}(E_\nu )\,X(\theta )/m_N\right]
\,,
\end{equation}
where 
$\rho$ is the density of the material in which the neutrino interaction occurs,  
$F_\nu = \sum_i (F_{\nu_i}+F_{\bar\nu_i})$ is the sum of the differential diffuse neutrino fluxes, 
$\sigma_{\nu N}^{\rm tot}$ denotes the total interaction cross section as a sum of the usual
Standard Model charged current cross section $\sigma_{\nu N}^{\rm cc}$ and the black hole production 
cross section~(\ref{sig_nuN_bh_pdf}),
\begin{equation}
\label{eq:ccpbh}
\sigma_{\nu N}^{\rm tot} = \sigma_{\nu N}^{\rm cc} + \sigma_{\nu N}^{\rm bh}\,,
\end{equation}
and, finally,  
$X(\theta  )$ is the column density of material between the detector and the upper atmosphere, as
a function of the zenith angle $\theta$. It can be approximated by~\cite{Morris:1991bb}
\begin{equation}
\label{eq:overburden}
X (\theta )  = \rho \biggl[
\sqrt{ (R_\oplus - D)^2 \cos^2\theta + 2DR_\oplus - D^2} -
(R_\oplus-D)\cos\theta
\biggr]\,, 
\end{equation}
where $R_\oplus$ denotes the radius of the earth and $D$ is the vertical depth 
of the detector. 

\begin{figure}[t]
\begin{center}
\includegraphics*[bbllx=20pt,bblly=221pt,bburx=570pt,bbury=608pt,%
width=12cm]{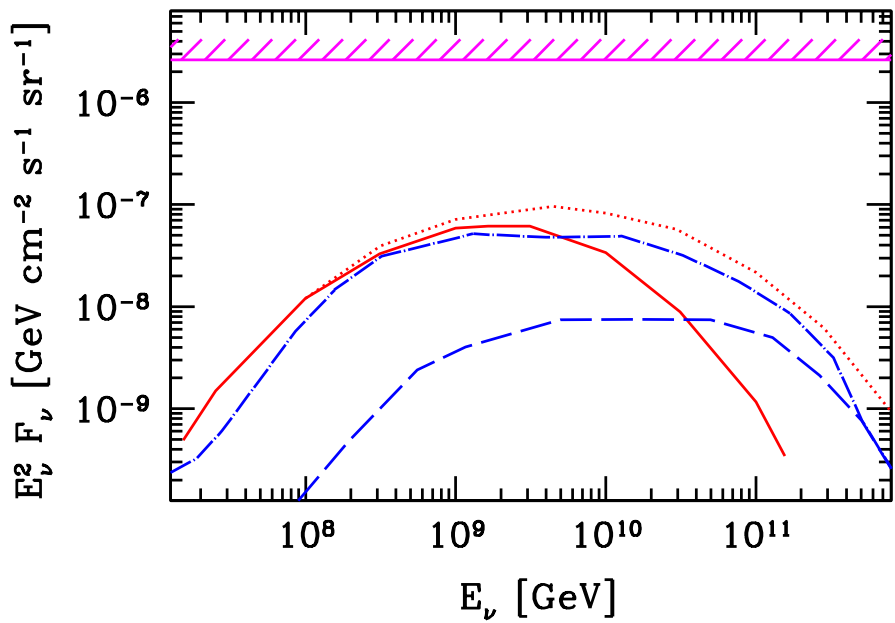}
\caption[dum]{\label{e2flux_cosmogenic}
Predictions of the cosmogenic neutrino flux, $F_\nu =\sum_i\left[ F_{\nu_i}+
F_{\bar\nu_i}\right]$. 
Long-dashed (long-dashed--dotted) line: Flux from 
Ref.~\cite{Yoshida:1993pt} for cosmological evolution parameters 
$m=2$, $z_{\rm max}=2$ ($m=4$, $z_{\rm max}=4$).
Solid (dotted) line: Flux from Ref.~\cite{Protheroe:1996ft}, 
assuming a maximum energy of $E_{\rm max}=3\cdot 10^{20(21)}$ eV for the ultrahigh energy cosmic rays.
Shaded: Theoretical upper limit of the neutrino flux from ``hidden'' 
astrophysical sources that are
non-transparent to ultrahigh energy nucleons~\cite{Mannheim:2001wp}.
}
\end{center}
\end{figure}

As in Ref.~\cite{Ringwald:2001vk}, we shall 
exploit both conservative lower and upper limits~\cite{Waxman:1999yy,Mannheim:2001wp} on the 
presently unknown differential flux $F_\nu$ of ultrahigh energy cosmic neutrinos  (for recent reviews, see  
Ref.~\cite{Protheroe:1999ei}) entering into
event rates at neutrino telescopes such as Eq.~(\ref{rate_contained}).
More or less guaranteed, and therefore comprising a reasonable lower bound, are the so-called 
cosmogenic neutrinos which are produced when ultrahigh energy cosmic protons
inelastically scatter off the cosmic microwave background radiation~\cite{Greisen:1966jv}
in processes such as $p\gamma\to \Delta\to n\pi^+$, where the produced pion subsequently 
decays~\cite{Beresinsky:1969qj}.
Recent estimates of these fluxes  can be found in 
Refs.~\cite{Yoshida:1993pt,Protheroe:1996ft,Yoshida:1997ie,Engel:2001hd}, some of which 
are shown in Fig.~\ref{e2flux_cosmogenic}. 
The upper limit from ``hidden'' hadronic astrophysical sources\footnote{For an early determination of such an upper limit, see 
Ref.~\cite{Berezinsky:1979pd}. Fluxes of this size are required in the context of the Z-burst 
scenario~\cite{Fargion:1999ft} for the highest
energy cosmic rays.}, i.\,e. from those sources from which only 
photons and neutrinos can escape, is much larger~\cite{Mannheim:2001wp} and  
also shown in Fig.~\ref{e2flux_cosmogenic}. 

\begin{figure}
\begin{center}
\includegraphics*[bbllx=20pt,bblly=221pt,bburx=585pt,bbury=608pt,%
width=13cm]{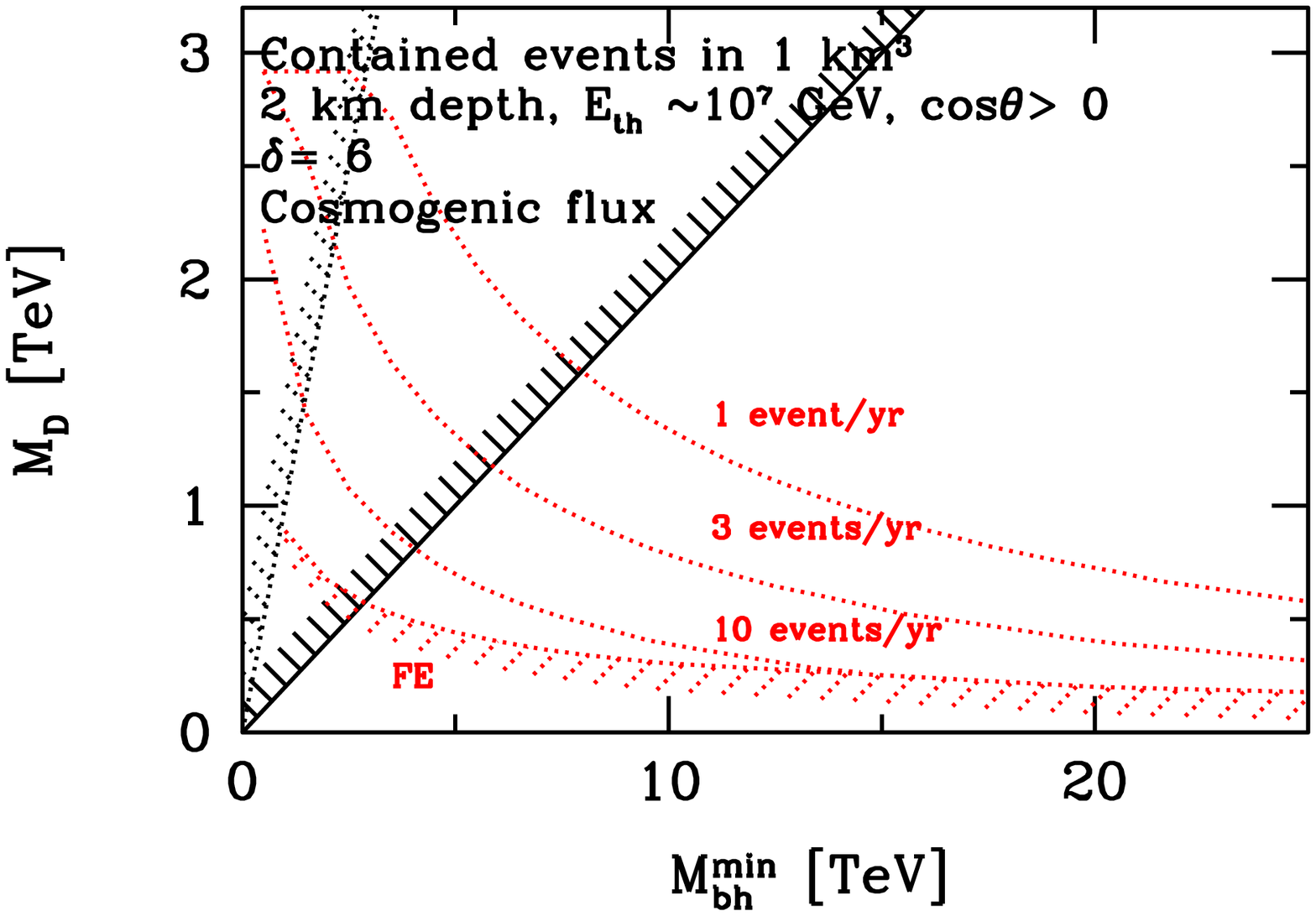}
\includegraphics*[bbllx=20pt,bblly=221pt,bburx=585pt,bbury=608pt,%
width=13cm]{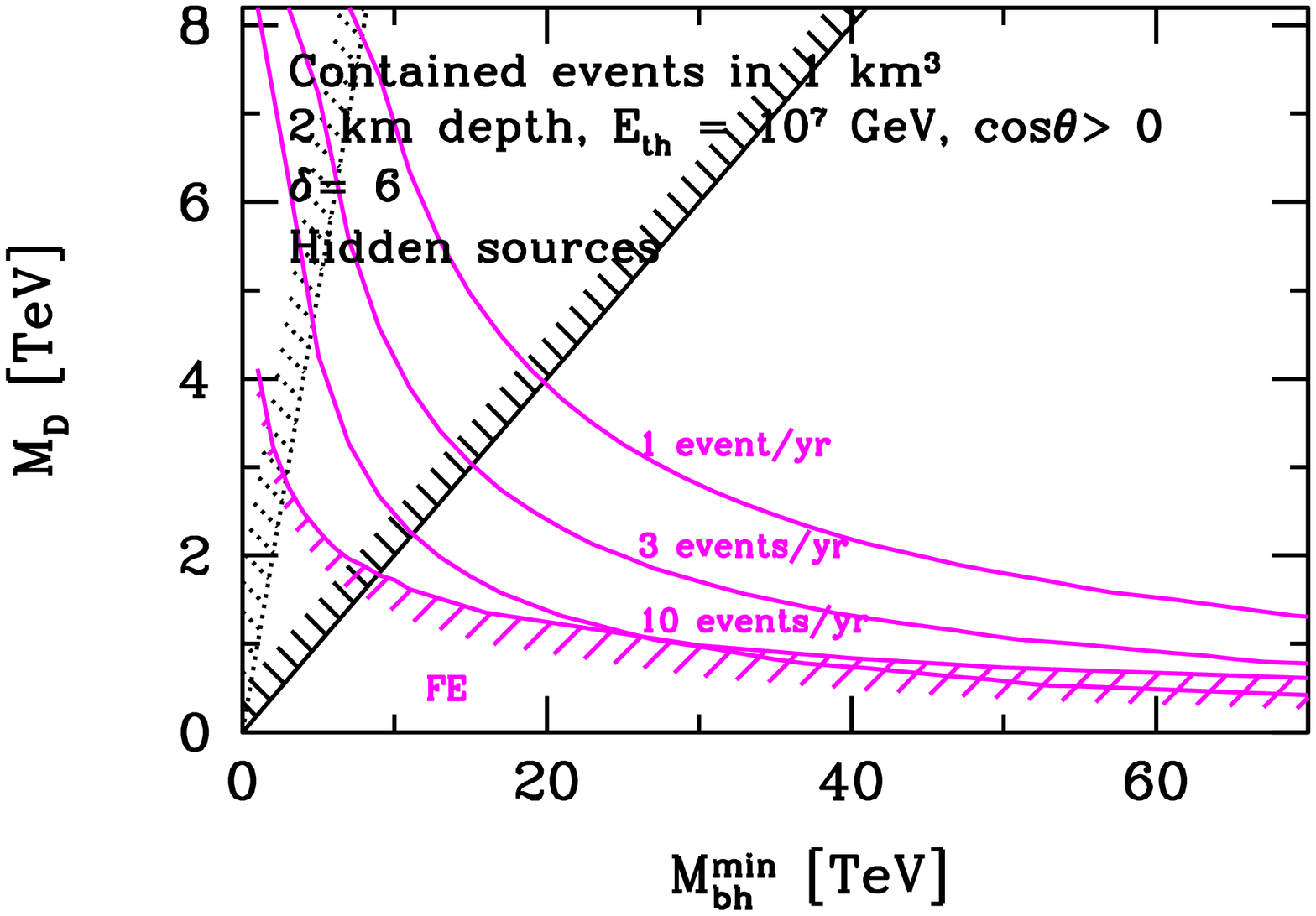}
\caption[dum]{\label{bh_par_cascades}
Reach in the black hole production parameters for $\delta = 6$ extra dimensions, 
for contained events in an under-ice detector at a depth of 2 km and with an 1 km$^3$ 
ficudial volume ($1\ {\rm yr}=3.16\cdot 10^7$~s). 
The shaded solid lines labeled ``FE'' indicates the constraint~\cite{Ringwald:2001vk} 
arising from the non-observation of 
horizontal showers by the Fly's Eye collaboration~\cite{Baltrusaitis:1985mt}.  
The shaded dotted, $M_D = M_{\rm bh}^{\rm min}$, and shaded solid, 
$M_D=(1/5)\,M_{\rm bh}^{\rm min}$, lines give a rough indication of the 
boundary of applicability of  the semiclassical picture~\cite{Giddings:2001bu}.
{\em Top:} Exploiting the 
cosmogenic neutrino flux from Ref.~\cite{Protheroe:1996ft} (cf. Fig.~\ref{e2flux_cosmogenic}).
{\em Bottom:} Exploiting 
the upper limit on the neutrino flux from ``hidden'' hadronic astrophysical sources from
Ref.~\cite{Mannheim:2001wp} (cf. Fig.~\ref{e2flux_cosmogenic}). 
}
\end{center}
\end{figure}

The projected reach in the black hole production parameter space for $\delta = 6$ extra dimensions 
for contained events in an under-ice detector corresponding to the IceCube proposal~\cite{Halzen:1999jy}
(2 km depth, fiducial volume 1 km$^3$) is shown in Fig.~\ref{bh_par_cascades}, for two of the
discussed ultrahigh energy cosmic neutrino fluxes. 
Table~\ref{tb:contained_sm} summarizes the corresponding expected Standard Model background rates 
for contained events due to charged current interactions.
Taking into account the small effective volume, $V\approx 0.001\div 0.01$ km$^3$, of the currently operating 
AMANDA and Baikal neutrino telescopes, it seems that, on the basis of contained events, these facilities
cannot compete with the already established limits from air shower observatories such as 
Fly's Eye and AGASA, and
the projected ones from Auger~\cite{Feng:2002ib,Ringwald:2001vk,Anchordoqui:2001cg}. 
Even IceCube does not seem to be really competitive, as far as the contained events are concerned: 
According to the time schedule of IceCube, the construction starts 
in 2004 and the final effective volume, $V\approx 1$ km$^3$, is reached only 
after the LHC starts operating and Auger has taken data for already a 
few years.
For the under-water/ice neutrino telescopes considered so far, the best perspective for detection of 
black hole events on the basis of the contained events might thus be assigned to 
RICE~\cite{Kravchenko:2001id}, a currently operating radio-Cherenkov telescope with an 
effective volume of about 1~km$^3$ for 10$^8$~GeV electromagnetic cascades~\cite{Seckel:2001}.

\begin{table}
\begin{center}
\begin{tabular}{|c||c|c|c|c|}
\hline 
 & \multicolumn{2}{|c|}{cosmogenic neutrinos} & \multicolumn{2}{|c|}{hidden sources}\\
\hline
$E_{\rm th}$& contained & through-going & contained & through-going \\
\hline    
$10^7$~GeV & 0.47   (yr km$^3)^{-1}$ & 0.85   (yr km$^2)^{-1}$ 
                      & 116  (yr km$^3)^{-1}$ & 92  (yr km$^2)^{-1}$\\
\hline
\end{tabular}
\caption{Event rates ($1\ {\rm yr}=3.16\cdot 10^7$~s) of contained reactions and 
through-going muons, respectively, for  
Standard Model like cross sections for the case of the cosmogenic neutrino flux from 
Ref.~\cite{Protheroe:1996ft}, assuming a maximum energy of $E_{\rm max}=3\cdot 10^{21}$ eV 
for the ultrahigh energy cosmic rays, and the case of the upper limit 
on the neutrino flux from ``hidden'' hadronic astrophysical sources from
Ref.~\cite{Mannheim:2001wp}, respectively. An under-ice detector at a depth of 2 km has
been assumed. For the contained events, we have considered,   
as in Eq.~(\ref{eq:ccpbh}), charged current reactions from all flavors, 
but have neglected neutral current reactions, because 
the fraction of visible energy, $\langle y\rangle\approx 0.2$, as well as 
the cross section, is lower for the latter~\cite{Gandhi:1996tf}.
For the through-going events, we have considered charged current $\nu_\mu N$
and $\bar\nu_\mu N$ interactions and an equipartion of neutrino flavors in the
neutrino flux.}
\label{tb:contained_sm}
\end{center}
\end{table}

{\em 4.}
The ability to detect muons from distant neutrino reactions increases an
under-ice/water detector's effective neutrino target volume dramatically
and is the premise upon which such detectors can act as neutrino telescopes. 
Let us therefore, in addition to the rate~(\ref{rate_contained}) of fully contained
black hole events, estimate the rate of through-going muons at a neutrino telescope. 
We define through-going muon events to be events with a vertex located 
outside of the detector and which then pass through the detector.
The expected rate for such events is then proportional to the area of
the detector. The muon rate above some threshold $E_{\rm th}$ per unit area $A$ and
solid angle $\Omega$ can be estimated by 
\begin{eqnarray}
\label{eq:trate}
\lefteqn{
\frac{{\rm d}^3 N^{\rm bh}_\mu}{{\rm d}A\,{\rm d}t\,{\rm d}\Omega} = \,
\sum_{k=1}^\infty \frac{1}{m_{ N}} 
\int_{E_{\rm th}}^\infty {\rm d}E_\nu\, F_\nu ( E_\nu )\, 
\int {\rm d}n_\mu\, \frac{{\rm d}\sigma^{\rm bh}_{\nu N}}{{\rm d}n_{\mu}}\ 
\frac{n_\mu^k}{k\,!}\,{\rm e}^{-n_\mu}
\,\times \,}
\\[1.5ex]\nonumber &&
\mbox{}\hspace{7ex}\times \int_{0}^{X(\theta )} {\rm d}X^{\prime} 
\exp\left[ -\sigma^{\rm tot}_{\nu N}(E_\nu )\,(X(\theta )-X^\prime )/m_N\right]
\left( 1- \left(1-\tilde{p}^{\rm bh}_\mu(E_\nu, \,E_{\rm th},X^{\prime})\right)^k\right)\,.
\end{eqnarray}
In words, the rate of through-going muon events from a black hole reaction is obtained by 
summing over different muon multiplicities $k$, weighted with the Poisson 
probability of mean $n_\mu$   
that $k$ muons emerge from black hole decay and that at least one can be observed in the detector.
The quantity $\tilde p_\mu^{\rm bh}$, which may be interpreted as the 
detection probability of a ``typical'' muon produced in a black hole
event, is defined as (cf. also Ref.~\cite{Morris:1991bb})
\begin{equation}
\label{eq:ptilde}
{\tilde p}_\mu^{\rm bh} (E_\nu , E_{\rm th}, X^\prime )
=
\int_{E_{\rm th}}^{E_\nu} {\rm d}E_\mu\,
\frac{1}{n_\mu^{\rm bh}}\frac{{\rm d}n_\mu^{\rm bh}}{{\rm d}E_\mu}\,
p_\mu (E_\mu , E_{\rm th}, X^\prime )\,
\end{equation}
with $p_\mu (E_\mu , E_{\rm th}, X^\prime )$ denoting the probability 
that a muon of definite energy $E_\mu$ traverses an amount $X^\prime$
of material with $E_\mu >E_{\rm th}$. 

\begin{figure}
\begin{center}
\includegraphics*[bbllx=20pt,bblly=221pt,bburx=585pt,bbury=608pt,%
width=13cm]{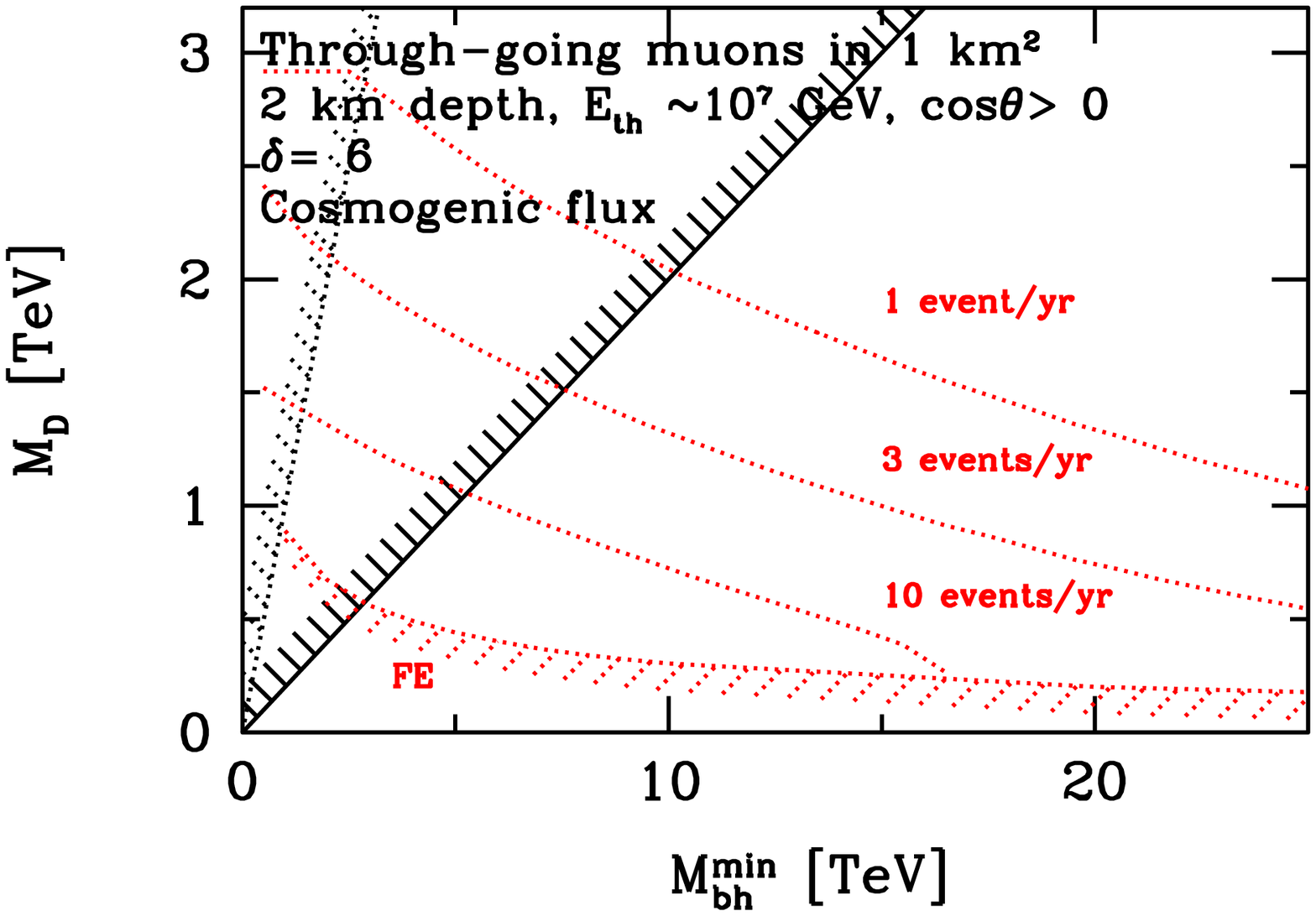}
\includegraphics*[bbllx=20pt,bblly=221pt,bburx=585pt,bbury=608pt,%
width=13cm]{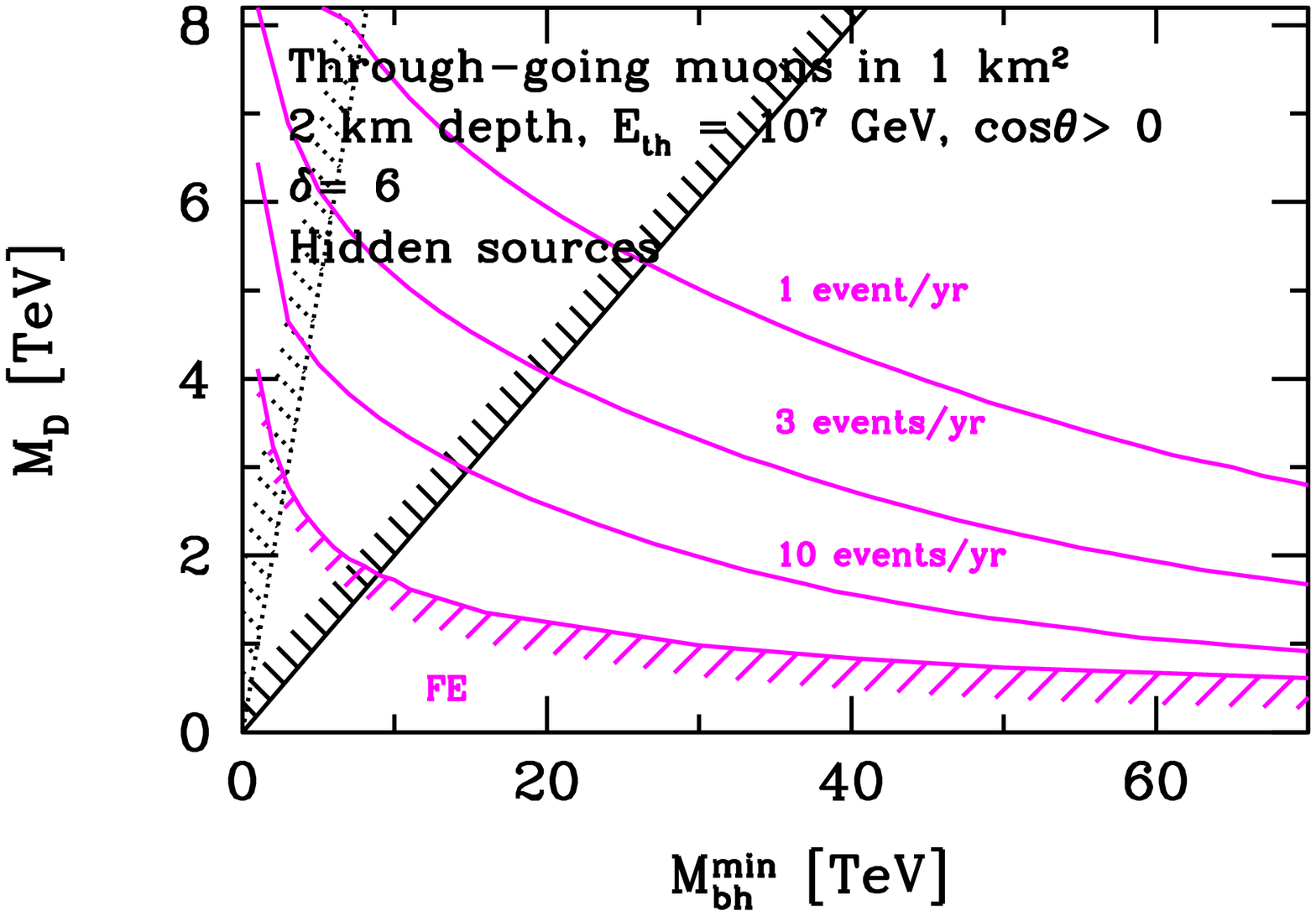}
\caption[dum]{\label{bh_par_through}
Reach in the black hole production parameters for $\delta = 6$ extra dimensions, 
for through-going muons in an under-ice detector at a depth of 2 km and with an 1 km$^2$ 
effective area ($1\ {\rm yr}=3.16\cdot 10^7$~s). 
The shaded solid lines labeled ``FE'' indicates the constraint~\cite{Ringwald:2001vk}  arising from the non-observation of 
horizontal showers by the Fly's Eye collaboration~\cite{Baltrusaitis:1985mt}.  
The shaded dotted, $M_D = M_{\rm bh}^{\rm min}$, and shaded solid, 
$M_D=(1/5)\,M_{\rm bh}^{\rm min}$, lines give a rough indication of the 
boundary of applicability of  the semiclassical picture~\cite{Giddings:2001bu}.
{\em Top:} Exploiting the 
cosmogenic neutrino flux from Ref.~\cite{Protheroe:1996ft} (cf. Fig.~\ref{e2flux_cosmogenic}).
{\em Bottom:} Exploiting 
the upper limit on the neutrino flux from ``hidden'' hadronic astrophysical sources from
Ref.~\cite{Mannheim:2001wp} (cf. Fig.~\ref{e2flux_cosmogenic}). 
}
\end{center}
\end{figure}

If one neglects the stochastic 
effects of range straggling which may become important for muon energies close
to the detection threshold,  i.e. within the approximation
of a deterministic energy-range relationship, one has the step 
function
\begin{equation}
p_\mu (E_\mu , E_{\rm th}, X^\prime ) = 
\theta (E_\mu - E_\mu^\prime (E_{\rm th},X^\prime ))\,
\end{equation}
where $E_\mu^\prime$ is the solution of 
\begin{equation}
X^\prime = \int_{E_{\rm th}}^{E_\mu^\prime}
\frac{{\rm d}E}{a (E) + b (E)\,E}\,.
\end{equation}
The ionization and radiation loss parameters,  $a (E)$ and $b (E)$, 
vary only slowly with energy and take the asymptotic values 
$a \approx 3\cdot 10^{-3}$ GeV cm$^2$/g and $b\approx 4\cdot 10^{-6}$ cm$^2$/g~\cite{Groom:2000in}.
Choosing these constant values, we get for the average muon range,
\begin{equation}
\label{eq:murange}
X^\prime \approx \frac{1}{b}\,
\ln\frac{a + b E^\prime_\mu}{ a + b E_{\rm th}}
\hspace{4ex}
\Leftrightarrow
\hspace{4ex}
E_\mu^\prime (E_{\rm th},X^\prime ) \approx 
\left[ E_{\rm th} + \left( 1 - {\rm e}^{- b\,X^\prime}\right)\,
\frac{a}{b}\right] {\rm e}^{b\,X^\prime}\,. 
\end{equation}
Since for our considerations $E_{\rm th}\sim 10^7$ GeV, we can safely neglect the contribution proportional to $a$ in the following. 

For the laboratory energy distribution of the muons produced in a 
black hole event, entering the probability~(\ref{eq:ptilde}), we have taken for 
simplicity\footnote{Alternatively, one may use a 
boosted thermal distribution, $\sim \int_{E_\mu/2\gamma}^\infty 
{\rm d}E (E/T_{\rm H})^2/(e^{E/T_{\rm H}}+1)$,  
which, however, needs an additional numerical integration.}
(cf. also Ref.~\cite{Morris:1991bb})
\begin{equation}
\label{eq:cmtolab}
\frac{1}{n_\mu^{\rm bh}}\frac{{\rm d}n_\mu^{\rm bh}}{{\rm d}E_\mu}
=\frac{\langle n\rangle}{2E_\nu}\,
\theta\left(\frac{2 E_\nu}{\langle n\rangle}-E_\mu \right)\,,
\end{equation}
with $\langle n\rangle$ being the average number~(\ref{eq:av_mult}) of particles produced in black hole decay 
(cf. Fig.~\ref{multiplicity}).
The distribution~(\ref{eq:cmtolab}) is obtained by boosting a delta-function energy spectrum, 
\begin{equation}
\label{eq:spec}
\frac{{\rm d}\sigma_{\nu N}^{\rm bh}}{{\rm d}E^{\rm cm}_\mu}
\propto \delta ( E_\mu^{\rm cm} - M_{\rm bh}/\langle n\rangle )\,,
\end{equation}
in the cm frame of the black hole, with isotropic direction of the decay particles, 
with a Lorentz factor $\gamma = E_\nu/M_{\rm bh}$ to the laboratory frame.  

Finally, we have exploited the ansatz (similar to Eq.(\ref{eq:spec})), 
\begin{equation}
\frac{d\sigma^{\rm bh}_{\nu N}}{{\rm d}n_\mu} 
= \sigma^{\rm bh}_{\nu N}\ \delta \left( n_\mu-\langle n_\mu \rangle\right)\,,
\hspace{6ex}
{\rm with\ }\ \langle n_\mu\rangle \approx \langle n\rangle /30 \,,
\end{equation}
to perform also the integration over the multiplicity $n_\mu$ in our rate
estimate~(\ref{eq:trate}).  The above relation for $\langle n_\mu \rangle$ is based on
the assumption that a black hole does not discriminate between Standard Model  particles in its 
evaporation spectrum~\cite{Giddings:2001bu}.

The projected reach in the black hole production parameter space for $\delta = 6$ extra dimensions 
for through-going muons in an under-ice detector at a depth of 2 km and with an
effective area of 1 km$^2$ is shown in Fig.~\ref{bh_par_through}, for two of the
discussed ultrahigh energy cosmic neutrino fluxes. 
The corresponding expected Standard Model background rates 
for through-going muons due to charged current interactions is summarized in 
Table~\ref{tb:contained_sm} (see also Ref.~\cite{Alvarez-Muniz:2001es} for related
calculations).
An additional source of background which 
is not related to an astrophysical neutrino flux is the muon flux arising 
from charm production by cosmic rays interacting in the atmosphere.
Using two available models~\cite{Bugaev:1998bi,Gelmini:2000ve} we have 
estimated a background of 
$0.33\div 1.5$ through-going muon events from atmospheric charm production per year.
To further reduce this background, one may introduce a cut in the zenith
angle, e.g. $\theta > 60^{\,o}$.

\noindent
{\em 5.} 
We have considered the sensitivity of neutrino telescopes to black hole production in the
context of extra dimension scenarios with TeV-scale gravity and compared it to
the one of air shower facilities and the LHC. 
For concreteness, we have assumed a detector with 1~km$^2$ area and 1~km$^3$ volume, 
respectively, located at a vertical depth of 2~km in ice.

We have seen that the effective volumina for contained events, $V\approx
0.001\div 0.01$~km$^{3}$, available from the currently operating AMANDA and
Baikal detectors  are too small 
to yield a large enough rate to be competitive with the already existing
constraints on black hole production from Fly's Eye and AGASA and with the 
projected ones from
Auger~\cite{Feng:2002ib,Ringwald:2001vk,Anchordoqui:2001cg}.  
The planned IceCube detector with a fiducial volume of about 1 km$^3$ 
would have an interesting perspective.
However, with the start of construction of IceCube in 2004 and full 
completion only after the physics start at the LHC, the discovery potential for
black hole production is strongly reduced.
Instead, the radio-Cherenkov telescope RICE~\cite{Kravchenko:2001id}, with an effective volume of about 
0.1~(1)~km$^3$ for 10$^{7(8)}$~GeV electromagnetic cascades~\cite{Seckel:2001}, could set sensible constraints using 
already available data.  

The prospects for a search for black hole signatures using through-going muons are also good. 
For a detection area of 1 km$^2$ we have shown that in case the neutrino flux 
is just on the level of the 
cosmogenic one, only a few ($\lwig 1$) contained or through-going background events from Standard Model charged
current interactions are expected per year. The background of prompt
muons from atmospheric charm was estimated to be of similar 
magnitude and can be further reduced by application of a cut on the 
zenith angle.
It was shown in Ref.~\cite{Hundertmark:2001}, that AMANDA achieves 
an effective area of about $0.3$ km$^2$ for down-going muons above an energy 
 of $10^{7}$~GeV and that suppression of the atmospheric muon background is possible.
With 5 years of AMANDA data available and an assumed duty-factor of $2/3$ 
this leads to an exposure of about 1 km$^2\cdot$yr. 
Therefore, strong constraints on the black hole production 
parameters are expected from AMANDA if the currently available data
show no through-going muons\footnote{It is noteworthy that in this case 
AMANDA can put also strong constraints on sphaleron production parameters~\cite{Morris:1991bb}, 
since the phenomenology of black hole and sphaleron production are quite similar.} above $10^7$~GeV. In  this case, 
the sensitivity of  AMANDA -- approximately the 3 events/yr contour in Fig.~\ref{bh_par_through} (top) for a 
90 \% confidence level limit --  
compares favorably with the one of Fly's Eye, AGASA, and Auger~\cite{Feng:2002ib,Ringwald:2001vk,Anchordoqui:2001cg}.
And even at the kinematic limit of the LHC, M$_{\rm bh}^{\rm min}\approx 10$~TeV (cf. Ref.~\cite{Ringwald:2001vk}),
one is still left with a sensitivity on $M_D$ of about $1.3$~TeV. 

In the case that the ultrahigh energy cosmic neutrino flux considerably exceeds the cosmogenic one, 
discrimination between Standard Model background and black hole events
becomes crucial. This might not be achieved by observation of through-going 
events alone, but with the additional information of contained events.
For example, the spectral information from contained and through-going events 
should be remarkably different, as the through-going muons from black hole decays 
are produced with considerably lower energy. 
Another discriminator may be based on the different hadronic components 
of Standard Model processes and black hole events. In a typical Standard Model charged 
current interaction one has an average hadronic energy fraction of 
$\langle y\rangle \approx 0.2$, whereas 
black hole events have typically a large hadronic component ($5:1$)~\cite{Giddings:2001bu} 
corresponding to a $\langle y\rangle \approx 0.8$. 
In particular radio-Cherenkov detectors such as RICE have the potential to 
discriminate between electromagnetic and hadronic showers by measuring
the characteristically different angular radiation patterns~\cite{Alvarez-Muniz:2000xx}. 

To summarize, we have shown that large scale under-water/ice neutrino detectors
have a substantial potential to discover or rule out black hole production which
is comparable to that of air shower facilities. In the optimistic case that a 
neutrino flux significantly higher than the cosmogenic one is realized in nature, 
one even has discovery potential for black holes at neutrino telescopes beyond the 
reach of LHC.

\section*{Acknowledgements}
We thank C. Spiering 
for fruitful discussions.

\end{document}